\documentclass[preprint]{elsarticle}

\usepackage{graphicx,hyperref, amsmath,amsfonts,amssymb}

\def\p{\partial}
\def\bn{\bgroup\bf n\egroup}
\def\x{{\tilde{x}}}
\def\y{{\tilde{y}}}
\def\z{{\tilde{z}}}
\def\vp{{\varphi}}
\def\ve{{\varepsilon}}
\def\const{\rm const}

\def\be{\begin{equation}}
\def\ee{\end{equation}} 
\journal{Physics Letters A}

\begin{document}
\begin{frontmatter}

\title{Thin nematic films: anchoring effects and stripe instability revisited}


\author[ovm]{O. V. Manyuhina}
\ead{oksanam\,@\,nordita.org}
\author[mba]{M. Ben Amar}

\address[ovm]{Nordita, KTH Royal Institute of Technology and Stockholm University,  Roslagstullsbacken~23, SE-10691 Stockholm, Sweden}

\address[mba]{Laboratoire de Physique Statistique, Ecole Normale Sup\'erieure, UPMC  Univ Paris~06, Universit\'e Paris Diderot, CNRS, 24 rue Lhomond, 75005  Paris,  France}

\begin{abstract}
We study theoretically the formation of long-wavelength  instability patterns observed at spreading  of nematic droplets on liquid substrates. The role of surface-like elastic terms such as saddle-splay and anchoring  in nematic films of submicron thickness is (re)examined by extending our previous work [Manyuhina et al {\it EPL},~{\bf 92}, 16005 (2010)] to hybrid aligned nematics. We identify the upper threshold for the formation of stripes  and compare our results with experimental observations. We find that the wavelength  and the amplitude of the in-plane director undulations can be related to the small but finite azimuthal anchoring.  Within a simplified model we analyse the possibility of non-planar base state below the Barbero--Barberi critical thickness.
\end{abstract}


\end{frontmatter}

\section{Introduction}

Thin nematic film spread on liquid substrate represents a fascinating system to study the influence of interfaces on the organisation of liquid crystal (LC) molecules. The pioneering works by Lavrentovich, Sparavigna and Pergamenshchik~\cite{LP:1994,LP:1995,sparav:1994,sparav:1995} were focused on understanding the spontaneous spreading of 5CB (4-pentyl-4'-cyanobiphenyl) LC  on glycerol. The authors observed the formation of spatially-periodic stripes in films of submicron thickness resulting from the competition between the surface and volume energy  namely between the  boundary conditions and the elastic bulk anisotropy of nematic LC. One of the first models~\cite{LP:1994}, accounting for these observations, suggests the  crucial role of the splay-bend elastic constant $K_{13}$, which should not vary more than by 5\,\% to fit the data. Another model~\cite{sparav:1994} quantifies the threshold thickness in agreement with experimental data,  while the wavelength of stripes  was not determined.

Recent systematic experiments performed by the group of Cazabat~\cite{ulysse:2008,ulysse:2009,ulysse:2009pc,ulysse:2010,cazabat:2011} suggest that the stripe phase occurs within the range of thickness $h_{c_1}\leqslant h\leqslant h_{c_2}$, where  the lower threshold $h_{c_1}\simeq 20$--30~nm and the upper threshold $h_{c_2}\simeq 0.5$--$0.6~\mu$m seem to be independent of the type of substrate and LC molecules. The period $L$ of the stripe domains  is controlled by the thickness $h$ of the film in a non-trivial way, yielding the ratio $L/h\propto 100$ for different LC systems~\cite{ulysse:2008}. The presence of the free nematic--air interface allows the thickness of the spreading  film to be adopted by the film itself  rather than being a fixed parameter. The striking feature of the experiments~\cite{ulysse:2008,ulysse:2009,ulysse:2009pc,ulysse:2010,cazabat:2011} is the existence of  a {\it forbidden range of thickness}, visible as a discontinuity in microdroplet profile. At the room temperature the lower boundary is a trilayer of molecules ($\xi_{LB}\approx3.5$~nm)  while the upper boundary is  $\xi_{UB}\approx20$--30~nm.  The structure of the film with $\xi_{UB}$ is unknown, and in fact  on water and glycerol striped domains coexist directly with trilayer and $\xi_{UB}$ is the thickness of the thinnest striped film $h_{c_1}$~\cite{ulysse:2009,ulysse:2010,cazabat:2011}. There is no theoretical approach which could account for these experimental observations.

Within the continuum theory of liquid crystals~\cite{degennes:book} nematics are characterised by the unit vector, called the director, specifying the averaged orientation of molecules, which tend to align parallel to each other. The presence of interfaces influences the surface ordering and orientation of nematic director, also known as anchoring,  which is far from being understood completely~\cite{jerome,osipov:1997,ulysse:2008}. The conventionally assumed degenerate azimuthal anchoring on liquid substrates, does not necessarily mean that LC molecules can align along any direction simultaneously without some energy cost. Rather it means that because of some microscopic (e.g. adsorption layer~\cite{zhuang:1995})  or geometric~\cite{lavrent:1992} reasons there is a tendency of molecules to align (spontaneously or not) along a common preferred direction on substrate interface. To describe the planar degenerate anchoring at the nematic--liquid substrate interface the fourth order terms of  the tensorial order parameter must be included in surface free energy~\cite{fournier:2005}. Although nematics are uniaxial in the bulk, the wetting layer may exhibit biaxiality due to the lower symmetry near the surface~\cite{fournier:2005}.  According to~\cite{biscari:2006} the biaxiality arises naturally close to a curved surfaces with imposed homeotropic anchoring (free interface with air), where the order tensor becomes biaxial along the principal directions of the surface. 

Our recent analysis~\cite{our:epl} shows that without accounting for azimuthal anchoring at either of the interface,  the wavelength  of stripes at the lower threshold is infinite, in contradiction with  experimental observations~\cite{sparav:1994,ulysse:2008,ulysse:2009pc,ulysse:2010,cazabat:2011}.  The presence of small azimuthal anchoring at one of the interfaces penalises strong in-plane distortions of the nematic director (stripe domains) and favours the common alignment of the molecules. However, contrary to the polar anchoring, the effective azimuthal anchoring can be thought of as an extrinsic parameter of our system, induced either  by the thickness gradient~\cite{sparav:1994,lavrent:1992,ulysse:2010}, biaxiality~\cite{fournier:2005,biscari:2006} or short-range intermolecular forces.  It is known~\cite{zhuang:1995} for example, that part of nCB molecule is buried in water with a preferential orientation dictated by its short-range interactions with water, resulting in adsorbed monolayer, which then leads to the azimuthal anisotropy in nCB's orientation distribution. 

In the present manuscript firstly  we generalise  the theoretical analysis~\cite{our:epl,barbero:2002} towards the distorted hybrid aligned nematic (HAN) state in presence of azimuthal anchoring at nematic--liquid substrate interface. This allows us to identify the {\it upper threshold} for stripe instability $h_{c_2}\simeq0.5~\mu$m as well as finite wavelength and amplitude of the perturbation. The predictions are robust, compatible with experimental observations of 5CB-7CB and MBBA on glycerol and water and do not depend on a particular choice of elastic constants, contrary to~\cite{LP:1994}.  Next we explore  the implications of simplified model on the director's configuration in unperturbed state by i)~introducing  an effective azimuthal anchoring  at both interfaces accounting for possible biaxial order~\cite{fournier:2005,biscari:2006} or ii)~assuming a certain scaling for the azimuthal and polar degrees of freedom of the nematic director. Within such an {\it ad hoc} approach we find twisted  state and modification of the Barbero--Barberi condition~\cite{BB:1983}. Clearly these approaches are oversimplified to understand the physical picture as a whole, where the interplay between hydrodynamic and elastic instabilitites~\cite{benamar:2001,cummings:2011,cummings:2012}, may lead to a  nontrivial film profile extending over the micrometers and thickness discontinuity, which occurs within nanometers scale~\cite{ulysse:2008,ulysse:2009,ulysse:2009pc,ulysse:2010,cazabat:2011}. Nevertheless, it is a good starting point to investigate the connections between molecular and mesoscopic scale phenomena quantitatively.

\section{Formulation of the problem\label{sect:energy}}

Nematic liquid crystals are described by the unit vector $\bn$, called the director, characterising the averaged orientation of molecules. The free energy  associated with distortions of this director in space is given by
\be\label{eq:fel}
{\cal F}_{el} =\frac 12\int_V dV \big\{K |\nabla \bn|^2 - K_{24}\nabla\cdot\big( \bn (\nabla \cdot\bn)+\bn \times \nabla\times \bn \big)\big\},
\ee
where we have assumed the one-constant approximation of the Oseen--Zocher--Frank free energy~\cite{degennes:book}, quadratic in the director derivatives. The elastic constant $K$ stands for the equal  splay, twist and bend elastic moduli~\cite{equality}, while the saddle-splay elastic constant $K_{24}$ should satisfy the Ericksen's inequalities~\cite{ericksen:1966}
\be\label{eq:ericks}
 |K_{24}|\leqslant K, \qquad K\geqslant0,
\ee
which guarantee the stability of the uniform nematic state $\bn=\const$.  The uniformly aligned director configuration can become unstable in presence of the  competing boundary conditions and the saddle-splay elastic term, favouring the distorted configuration of the nematic director. Therefore the modulated stripe state can be achieved, when the saddle-splay contribution to the free energy becomes substantial, which is plausibly realised in {\it thin} nematic films with {\it weak} anchoring boundary conditions. 

For convenience  we  parametrise the director as
\be \label{eq:bn}
\bn= \sin\theta\cos\vp\, {\bf e}_x+ \sin\theta\sin\vp \, {\bf e}_y + \cos\theta\, {\bf e}_z,
\ee
where $\theta$ and $\vp$ are the polar and the azimuthal angles, depending on the Cartesian coordinates $x,y,z$.  
The first term of the free energy~\eqref{eq:fel} integrated over the volume $V$ characterises the bulk contribution and within the parametrisation \eqref{eq:bn} can be rewritten as 
\be\label{eq:fb}
{\cal F}_b= \frac K2 \int_{V } dV\,\big(|\nabla \theta|^2+\sin^2\theta|\nabla \vp|^2\big),
\ee
while the last divergence term can be transformed into the surface integral as
\be\label{eq:fs}
{\cal F}_s= -\frac {K_{24}}2 \int_{S } dS\,\big[\cos\vp(\p_x\theta +\cos\theta\sin\theta\p_y \vp)  +\sin\vp(\p_y\theta -\cos\theta\sin\theta\p_x \vp)\big]\bigg|_{z=0}^{z=h}.
\ee
where $h$ is the thickness of the nematic film and  the surface normal in flat central part of the spread film is assumed to be  parallel to the $z$-axis. 

We assume that the anchoring energy admits the general Rapini--Papoular form~\cite{rapini:1969}
\be\label{eq:fa}
{\cal F}_{a}=\frac 12\sum_{i=1,2}\int_{S} dS\,\big \{W_{\theta_i} \sin^2(\theta_i-\bar\theta_i)+W_{\varphi_i}\sin^2\theta_i\sin^2(\varphi_i-\bar\varphi_i)\big\},
\ee
where the integration is done over  the nematic liquid substrate interface $i=1$ ($z=0$) and the nematic--air interface $i=2$ ($z=h$). Here $W_{\theta_i}$ is the polar anchoring strength, accounting for energy cost due to the deviation of the polar angle  $\theta_i$ ($\theta_1\equiv \theta(0)$ and $\theta_2\equiv \theta(h)$) from the easy axis  $\bar\theta_1=\pi/2$ (planar anchoring)  and  $\bar\theta_2=0$ (homeotropic anchoring), respectively.   The second term  in ${\cal F}_a$ accounts for the deviation of the director's projection on the $xy$-plane from  a common alignment, which happens during the formation of stripes. From the experimental works~\cite{sparav:1994,ulysse:2009pc}  we know  that  the planar anchoring $W_{\theta_1}$ is stronger than the homeotropic anchoring $W_{\theta_2}$, and quantitatively the extrapolation lengths for 5CB on glycerol system are $L_{\theta_2}=K/W_{\theta_2} = 0.7~\mu{\rm m}$ and $L_{\theta_1}=K/W_{\theta_1}=0.35~\mu{\rm m}$.  According to~\cite{jerome} the value of the azimuthal anchoring $W_{\varphi_{1,2}}$ should be one or two orders of magnitude smaller than the polar anchoring, but in general it is  not fixed {\it a priori}.

In the following sections we assume the vanishing azimuthal anchoring at the nematic--air interface $W_{\vp_2}=0$, while $W_{\vp_1}\neq0$, and explore the predictions of the formulated model from the bifurcation analysis point of view. In section~\ref{ssect:wf2} we extend theoretical analysis towards the alternative case, when  $W_{\vp_2}\neq0$ and indicate the consequences of the proposed model for the director's configuration, without referring  to any particular experiment.

\section{Linear stability analysis\label{sect:han}}

\begin{figure}[t]
\centering
\includegraphics[width=0.8\linewidth]{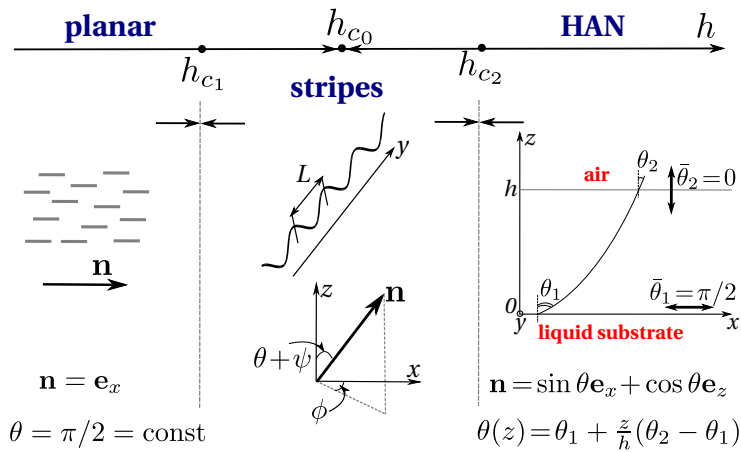}
\caption{Schematic representation of the structural phase transitions between the planar state, striped state ($h_{c_1} < h < h_{c_2}$) with periodicity $L$ in the $y$-direction, and the hybrid aligned nematic, where the film thickness $h$ is thought of as an order parameter.  The critical thickness between the planar and HAN state is  $h_{c_0}=L_{\theta_2}-L_{\theta_1}$~\cite{BB:1983}.}
\label{fig:stripe}
\end{figure}

In this section we assume the vanishing azimuthal anchoring at the nematic--air interface $W_{\vp_2}=0$, yielding base states with $\vp\equiv\bar\varphi_1=\const$, confined to the 2D plane. Hence, the angle $\theta(\z)$  depends only on the coordinate along the thickness of the film $\z=z/h$ and satisfies the Euler--Lagrange equation
\be\label{eq:theta}
\p_{\z\z} \theta=0, \quad \mbox{yielding}\quad \theta(\z)=\theta_1+(\theta_2-\theta_1)\z.
\ee
The equilibrium angles $\theta_i$, between the director $\bn$ and the $z$-axis at liquid substrate $\z=0$  and at air $\z=1$  interfaces  should satisfy the following boundary conditions for a given thickness $h$
\begin{subequations}\label{eq:bc}
\begin{align}
2(\theta_1-\theta_2)-\frac h {L_{\theta_1}}\sin2\theta_1&=0, \\
2(\theta_1-\theta_2)-\frac h {L_{\theta_2}} \sin2\theta_2&=0.
\end{align}
\end{subequations}
We will distinguish two base states 
\begin{itemize}
\item the homogeneous planar state with $\theta=\theta_1=\theta_2\equiv\pi/2$ ($h<h_{c_0}$),
\item  the HAN  state with $\theta_1\neq\theta_2$ ($h>h_{c_0}$) satisfying \eqref{eq:bc} with $\theta$~\eqref{eq:theta}.
\end{itemize}
The  critical thickness $h_{c_0}\equiv L_{\theta_2}-L_{\theta_1}$ (Barbero--Barberi~\cite{BB:1983}) characterises the transition between the planar and the HAN states. To find the critical thresholds  for the formation of stripes we consider the variation of the total free energy ${\cal F}={\cal F}_b +{\cal F}_s +{\cal F}_{a}$ given by the equations \eqref{eq:fb}--\eqref{eq:fa} with respect to the small perturbation of the director $\bn$~\eqref{eq:bn}. The  perturbation angles $\psi$ (polar) and $\phi$ (azimuthal) are assumed to be  small $O(\varepsilon)$ with periodic modulation in $y$-direction 
\be\label{eq:psi}
\psi(\y,\z)=\ve f(\z)\sin (\chi \y),\quad \phi(\y,\z)=\ve g(\z)\cos (\chi \y),
\ee
where  $\y=y/h$ is the dimensionless coordinate, and  $\chi=2\pi h/L$ is the wave\-number,  with $L$ being the period of stripes. 
Close to the instability threshold we can introduce a small parameter $\ve$, defined as $h=h_c(1+\ve^2)$, where $h_c$ is the critical thickness, characterising the instability of planar (HAN) states towards the stripe state. 
The difference in the total free energy of the stripe state and the base state can be written as the Taylor series  $\delta {\cal F}=\ve^2\delta{\cal F}^{(2)}+\ve^4\delta {\cal F}^{(4)}+\ldots$, with
\begin{multline}\label{eq:dF2}	
\delta{\cal F}^{(2)}=\frac {K\chi} {4\pi h}\int_0^{\frac {2\pi}\chi}\!\!\! d\y\bigg[\int_0^1\! \!d\z \,\big\{\overbrace{ \sin^2\theta(\z)\big [(\p_\z\phi)^2+(\p_\y\phi)^2\big ]+(\p_\z\psi)^2+(\p_\y\psi)^2}^{\omega_b^{(2)}} \big\}+ \\+\underbrace{\frac{h \sin^2\theta_1}{L_{\vp_1}} \phi(0)^2-\frac{h \cos2\theta_1}{L_{\theta_1}}\psi(0)^2
+\frac{h \cos2\theta_2}{L_{\theta_2}}\psi(1)^2 +2\vartheta\sin^2\theta\,\psi\,\p_\y\phi\bigg|_{\z=0}^{\z=1}}_{\omega_s^{(2)}}\bigg],
\end{multline}
where $\vartheta= K_{24}/K$ is another dimensionless parameter related to $p=(K_{22}+K_{24})/K$ in~\cite{our:epl} as  $\vartheta=p-1$. The value of $K_{24}$ as well as the extrapolation length for azimuthal anchoring $L_{\vp_1}$ are not precisely known, therefore we present below our results for the critical threshold as function of~$\vartheta$ and~$L_{\vp_1}$.

The Euler--Lagrange equations associated with \eqref{eq:dF2}  are
\begin{subequations}
\label{eq:ELhan}
\begin{align}
\p_{\z\z}f-\chi^2 f &=0,\\
\p_\z\big(\sin^2\theta \p_\z g\big) -\chi^2\sin^2\theta\, g&= 0,
\end{align}
\end{subequations}
with the general solution given by:
\begin{subequations}
\label{eq:fg}
\begin{align}
f(\z)&=C_1e^{\chi \z}  + C_2 e^{-\chi\z},\\
g(\z)&=\frac{C_3 e^{ \lambda\z} +C_4e^{-\lambda\z}}{\sin\theta(\z)},
\end{align}
\end{subequations}
where $\theta(\z)=\theta_1+(\theta_2-\theta_1)\z$ and $\lambda=\sqrt{\chi^2-(\theta_2-\theta_1)^2}$. In the planar case $\theta=\pi/2$  equation~\eqref{eq:fg}b simplifies to the sum of two exponents. The integration constants $C_i$ can be found from the boundary conditions associated with \eqref{eq:dF2}
\begin{subequations}\label{eq:bc2}
\begin{gather}
\bigg(-\frac{\p \omega_b^{(2)}}{\p \psi_{,\z}}+\frac{\p \omega_s^{(2)}}{\p \psi}\bigg)\bigg|_{\z=0}=0,\quad
\bigg(-\frac{\p \omega_b^{(2)}}{\p \phi_{,\z}}-\frac d{d\y}\frac{\p \omega_s^{(2)}}{\p \phi_{,\y}}+\frac{\p \omega_s^{(2)}}{\p \phi}\bigg)\bigg|_{\z=0}=0,\\
\bigg(\frac{\p \omega_b^{(2)}}{\p \psi_{,\z}}+\frac{\p \omega_s^{(2)}}{\p \psi}\bigg)\bigg|_{\z=1}=0,\quad
\bigg(\frac{\p \omega_b^{(2)}}{\p \phi_{,\z}}-\frac d{d\y}\frac{\p \omega_s^{(2)}}{\p \phi_{,\y}}+\frac{\p \omega_s^{(2)})}{\p \phi}\bigg)\bigg|_{\z=1}=0,
\end{gather}
\end{subequations}
which can be cast into the matrix form  as $\sum_{i=1}^4{\cal M}_{ij}C_i=0$, where
\begin{equation}\label{eq:matrix}
{\cal M}=
\begin{pmatrix}
 -\chi-\frac{h \cos2 \theta_1}{L_{\theta_1}} & \chi -\frac{h \cos2 \theta_1}{L_{\theta_1}} & \vartheta \chi  \sin\theta_1 & \vartheta \chi  \sin \theta_1 \\
 \frac{e^{\chi } (L_{\theta_2} \chi +h \cos2 \theta_2)}{L_{\theta_2}} & \frac{e^{-\chi } (h \cos2 \theta_2 -L_{\theta_2} \chi )}{L_{\theta_2}} & {-\vartheta e^{\lambda } \chi  \sin\theta_2} & -\vartheta e^{-\lambda}\chi  \sin\theta_2 \\
 \vartheta \chi  & \vartheta \chi  & \frac{h-L_{\vp_1} \lambda +L_{\vp_1} (\theta_2-\theta_1) \cot\theta_1}{ \sin\theta_1 L_{\vp_1}  } & \frac{h+L_{\vp_1} \lambda +L_{\vp_1} (\theta_2-\theta_1) \cot\theta_1}{ \sin\theta_1L_{\vp_1}} \\[1ex]
-\vartheta\chi e^{\chi }  & -\vartheta \chi e^{-\chi }   & \frac{e^{\lambda } (\lambda -(\theta_2-\theta_1) \cot\theta_2) }{\sin\theta_2 } & -\frac{e^{-\lambda } (\lambda +(\theta_2-\theta_1) \cot\theta_2)}{\sin\theta_2}
\end{pmatrix}.
\end{equation}


\begin{figure}[thb]
\centering
\includegraphics[width=0.75\linewidth]{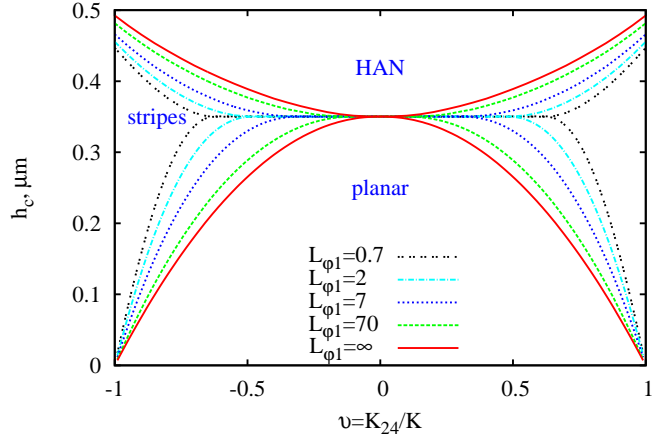}
\caption{The critical thickness, characterising the lower ($h_{c_1}$) and the upper ($h_{c_2}$) instability thresholds  as function of saddle--splay elastic constant $\vartheta=K_{24}/K$. The stripe phase exists between the curves of the same colour and it is restrained when $L_{\vp_1}$ decreases or azimuthal anchoring at nematic--liquid substrate interface  $W_{\vp_1}$ increases. The values for the polar anchoring are known $L_{\theta_1}=0.35~\mu$m and $L_{\theta_2}=0.7~\mu$m~\cite{ulysse:2009pc,ulysse:thesis}.}
\label{fig:hc}
\end{figure}

\begin{figure}[thb]
\centering
\includegraphics[width=0.75\linewidth]{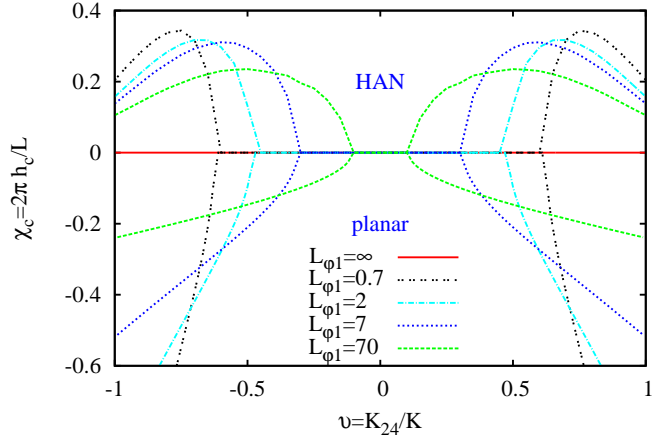}
\caption{The critical wavenumber $\chi_{c_1}$ and $\chi_{c_2}$, characterising the lower  and the upper instability thresholds, respectively. In the limit of the vanishing azimuthal anchoring $L_{\vp_1}\to \infty$ the wavelength of stripes vanishes.}
\label{fig:xc}
\end{figure}

The nontrivial solution ($C_i\neq0$) exists if and only if the determinant of the matrix~\eqref{eq:matrix} is zero, $\det{\cal M}=0$. This condition gives the relationship between  the thickness $h$ of the nematic film, the dimensionless wavenumber~$\chi$, the angles $\theta_{1,2}$ (HAN case) and other parameters of the system. Since the critical thickness of the nematic film $h_c$ is unknown, the problem should be solved consistently together with the boundary conditions for the equilibrium angles $\theta_i$ \eqref{eq:bc}, which in turn also depend on~$h$.  The minimum of the curves $\det{\cal M}=0$  in the  $h$--$\chi$ plane allows to identify the critical thickness $h_c$ and the critical wavenumber $\chi_c$ at the bifurcation point, for a given value of $\vartheta$, and the extrapolation length of azimuthal anchoring $L_{\vp_1}$, while the polar anchoring is known $L_{\theta_1}=0.35~\mu$m, $L_{\theta_2}=0.7~\mu$m~\cite{ulysse:2009pc,ulysse:thesis}. In Fig.~\ref{fig:hc} we plot $h_c$, characterising the planar--stripe ($h_{c_1}$)  and the HAN--stripe ($h_{c_2}$) instability thresholds. The curves are symmetric around  $\vartheta=0$. In presence of strong azimuthal anchoring (small values of $L_{\vp_1}$) the formation of a stripe phase is suppressed, as was predicted in~\cite{sparav:1991}.   Experimentally compatible values for critical thickness $h_{c_1}\simeq 40$~nm and $h_{c_2}\simeq 0.5~\mu$m occur at $|\vartheta|\simeq 1$ and for $L_{\vp_1}$ of order of tens of micron, corresponding to extremely weak azimuthal anchoring.

\begin{figure}[thb]
\centering
\includegraphics[width=0.75\linewidth]{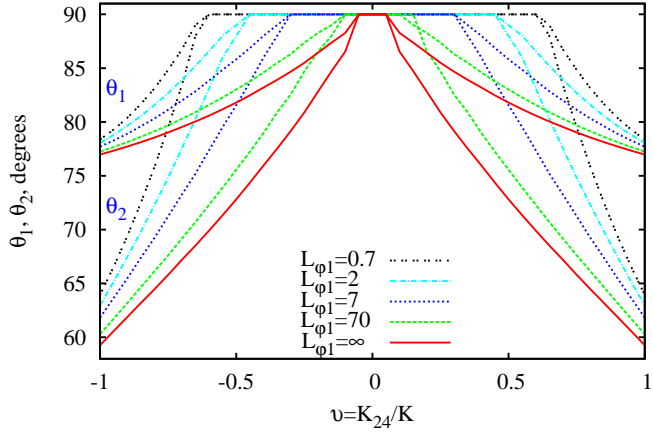}
\caption{The equilibrium angles $\theta_{1,2}$, satisfying the boundary conditions \eqref{eq:bc} (for $\vp=0$) at the critical thickness $h_{c_2}$ (see Fig.~\ref{fig:hc}). The values for the polar anchoring are known $L_{\theta_1}=0.35~\mu$m and $L_{\theta_2}=0.7~\mu$m~\cite{ulysse:2009pc,ulysse:thesis}.}
\label{fig:angle}
\end{figure}

In Fig.~\ref{fig:xc} we plot the critical wavenumber $\chi_{c_i}$ at $h_{c_i}$, with one inverse branch for illustrative purposes. As mentioned above, the experimentally relevant critical thickness $h_{c_i}$ occurs at $K_{24}\simeq K$ and $L_{\vp_1}\simeq 70~\mu$m, yielding the critical wavenumbers $\chi_{c_1}\simeq 0.2$ and $\chi_{c_2}\simeq 0.1$.  For planar--stripe instability $\chi$ vanishes only in the limit of $L_{\vp_1}\to \infty$ ($W_{\vp_1}=0$), while for HAN--stripe instability $\chi=0$ is always the solution of $\det{\cal M}=0$. Comparing the results on $h_{c_1}$ and $\chi_{c_1}$ with our previous analysis~\cite{our:epl}, where we have taken into account the  twist elastic constant $K_{22}\neq K_{11}$ and assumed  a non-zero azimuthal anchoring at nematic--air interface, we conclude that the predictions of both critical parameters agree quantitatively. This plausibly means that the ratio $K_{22}/K_{11}$ does not  play a dominant role in our system and that the one constant approximation is reasonable to analyse the  stripe instability. It is hard to directly compare theoretical predictions of the upper threshold $h_{c_2}$ and $\chi_{c_2}$ with experimental data, because of the presence of defects in nematic films  thicker than $h\simeq 0.5~\mu$m~\cite{ulysse:2008,ulysse:2010,cazabat:2011}. The   HAN state, therefore, is an idealised director's configuration which is probably not accessible in real systems, because of the presence of defects.  For completeness in Fig.~\ref{fig:angle}   we plot the equilibrium angles $\theta_{1,2}$ found at~$h_{c_2}$ and satisfying the  boundary conditions~\eqref{eq:bc}.

\section{Weakly nonlinear analysis\label{sect:ampl}}

\begin{figure}[tb]
\centering
\includegraphics[height=50mm]{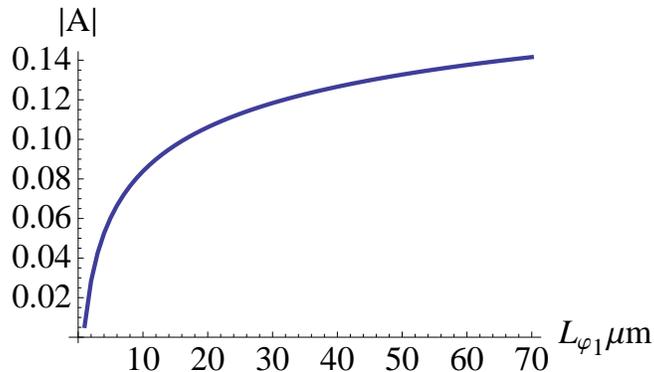}
\caption{The amplitude for the planar--stripe instability  as function of the azimuthal anchoring $L_{\vp_1}$.  The choice of parameters is  $L_{\theta_1}=0.35~\mu$m, $L_{\theta_2}=0.7~\mu$m, $|\vartheta|=0.95$,  $L_{\vp_2}=0$ and the corresponding thresholds are found in section~\ref{sect:han}.}
\label{fig:Aplan}
\end{figure}

The linear stability analysis, considered above allows to identify the critical thickness and the critical wavenumber at the instability threshold. However the linearised boundary conditions \eqref{eq:matrix} give only the ratio of the integration constants $C_i$. Thus, the perturbation functions $\phi$ and $\psi$  are {\it known} up to some multiplying factor  $A$, called the  amplitude of perturbation.  To eliminate this uncertainty and to find $A$ we expand the total free energy further in powers of $\ve$ up to $O(\ve^4)$, fourth order in $\phi$ and $\psi$. Moreover, we assume to be in the vicinity of the threshold $h_c,\chi_c$,  and the thickness of the film given by  $h=h_c(1+\varepsilon^2)$ is the control parameter of our system, with  $\varepsilon\ll1$  measuring the distance from the threshold. Then the Taylor expansion of the free energy is
\be\label{eq:expand}
\delta{\cal F}=\ve^2\, A^2\delta {\cal F}^{(2)}+\ve^4\,\big(A^2\delta {\cal F}_a^{(2)} +A^4\delta {\cal F}^{(4)}\big) + O(\ve^6),
\ee
where $\delta {\cal F}^{(2)}$ is given by \eqref{eq:dF2} for planar (HAN) case,  $\delta{\cal F}_a^{(2)}$ is the second order contribution of the anchoring energy and $\delta{\cal F}^{(4)}$ is the fourth order contribution of the total energy, expanded in $\phi$ and $\psi$.  We make use of {\it Mathematica} to integrate directly \eqref{eq:expand}, since the functions $\phi$ and $\psi$~\eqref{eq:fg} as well as the critical point were found before in section~\ref{sect:han}. The first term vanishes at the critical point $\delta{\cal F}^{(2)}\big|_{\substack{h=h_c,\\\chi=\chi_c}}=0$, as expected. Then the amplitude can be identified by extremising the next order term, namely
$A^2=-\delta{\cal F}_a^{(2)}/(2\delta{\cal F}^{(4)})\Big|_{\substack{h=h_c,\\\chi=\chi_c}}$. As pointed out in~\cite{gaetano:2008} this procedure may be equivalent to the derivation of the amplitude equations. The energy method, however, does not allow to find a nonlinear change in the wavelength. 

In Fig.~\ref{fig:Aplan} we plot the amplitude $A$ for the planar--stripe instability as function of  the azimuthal anchoring $L_{\vp_1}$. The increase of $A$ with $L_{\vp_1}$ suggests the suppression of the stripe phase in presence of strong azimuthal anchoring, which  also agrees with the linear stability analysis performed in section~\ref{sect:han}.  For all considered values of $L_{\vp_1}$ the amplitude is $O(1)$, justifying the perturbation approach developed above.  

The analysis developed above predicts the threshold for stripe instability, compatible with experimental observations, the question about the base state, however, remains a sensitive issue of the model as mentioned above in the introduction. The assumption about `idealised' base states, such as planar and HAN, is not strictly supported by recent experiments, since  the structure of the film with 20--30~nm thickness is not known while at the upper threshold  $\approx 0.5$--$0.6~\mu$m stripes are replaced by other complex structures with defects~\cite{ulysse:2008,ulysse:2010,cazabat:2011}.  The alternative 2D model for the spreading of nematic drops accounting for defects and the evolution of profile was proposed in~\cite{cummings:2011,cummings:2012}, where  an {\it ad hoc} anchoring condition which relaxes anchoring strength when thickness of the film goes to zero was introduced. Indeed,  the Rapini--Papoular anchoring condition is plausibly not suitable for an ultrathin nematic films with a free interface.

\section{On the base state of  nematic films\label{sect:base}}

In this section, without referring to any particular experiment, we explore the possibility of different director configurations along the thickness of the film i) by imposing an effective azimuthal anchoring at both interfaces~\eqref{eq:fa} ($W_{\vp_2}\neq0$), ii) by assuming a certain scaling of the in-plane and the out-of-plane degrees of freedom, leading to their coupling. Then within the developed {\it ansatz} we look for the critical thickness between the planar state and distorted state, yielding the modified Barbero--Barberi condition $h_{c_0}=L_{\theta_2}-L_{\theta_1}$~\cite{BB:1983}.

\subsection{In presence of azimuthal anchoring $W_{\vp_2}\neq0$\label{ssect:wf2}}

The presence of effective azimuthal anchoring  might account for biaxiality of nematic order close to the surface~\cite{fournier:2005,biscari:2006}, thickness gradient~\cite{sparav:1994,lavrent:1992,ulysse:2010} or short-range interactions~\cite{zhuang:1995,osipov:1997}. 
We make use of the continuous approach formulated above in the section~\ref{sect:energy}. For flat films the angles $\theta(\z)$ and $\vp(\z)$~\eqref{eq:bn} depend  only on the coordinate $\z=z/h$ along the thickness of the film and the Euler--Lagrange equations associated with the bulk free energy \eqref{eq:fb}  are 
\begin{subequations}
\label{eq:EL}
\begin{align}
\p_{\z\z}\theta-\sin\theta\cos\theta (\p_\z\vp)^2&=0,\\
\p_\z(\sin^2\theta \p_\z\vp)&=0.
\end{align}
\end{subequations}
The first integral is
\be\label{eq:EL1}
\p_\z\vp=\frac {B_1}{\sin^2\theta},\qquad (\p_\z\theta)^2=B_2^2-\frac{B_1^2}{\sin^2\theta},
\ee
and the second one yields the solutions $\theta$ and $\vp$, written in the following form
\be\label{eq:newgs}
\cos\theta(\z) =\frac{\sqrt{B_2^2-B_1^2}}{B_2}\sin(B_3-B_2\z),\quad \tan(\vp(\z)+B_4)=\frac{B_1}{B_2}\tan(B_2\z-B_3).
\ee
Let us consider several limiting cases: 
\begin{itemize}
\item[a)] $\theta_1=\theta_2=\pi/2$ and $\vp_1=\vp_2$, ($B_1=0$, $B_2=0$), {\bf planar} state with uniformly aligned director  $\bn$,\\[-1ex]
\item[b)] $\theta(\z)= \theta_1+(\theta_2-\theta_1)\z$ and $\vp_1=\vp_2$, ($B_1=0$), {\bf HAN} state with $\theta$ varying along the thickness of the film\\[-1ex] 
\item[c)] $\theta_1=\theta_2=\pi/2$ and  $\vp(\z)=\vp_1+(\vp_2-\vp_1)\z$, ($B_2^2=B_1^2$), {\bf twisted} state with $\vp$ varying along the thickness,\\[-1ex]
\item[d)] $\theta_1\neq\theta_2$ and  $\vp_1\neq\vp_2$, satisfying  \eqref{eq:newgs}, \eqref{eq:4bc}, {\bf twisted--bent}.
\end{itemize}

In general, the integration constants $B_i$ should be determined from the boundary conditions associated with \eqref{eq:fb},  \eqref{eq:fa} expressed as
\begin{subequations}
\label{eq:4bc}
\begin{align}
2\p_\z\theta \big|_{\z=0}+ \sin2\theta_1\bigg(\frac h{L_{\theta_1}}-\frac h{L_{\vp_1}} \sin^2\vp_1\bigg)&=0, \\
2\p_\z\theta \big|_{\z=1}+ \sin2\theta_2\bigg(\frac h{L_{\theta_2}}+\frac h{L_{\vp_2}} \sin^2(\vp_2-\bar\varphi)\bigg)&=0, \\
\sin^2\theta_1\bigg(2\p_\z\vp \big|_{\z=0}-\frac h{L_{\vp_1}} \sin2\vp_1\bigg)&=0, \\
\sin^2\theta_2 \bigg(2\p_\z\vp \big|_{\z=1}+\frac h{L_{\vp_2}} \sin^2\theta_2\sin2(\vp_2-\bar\varphi)\bigg)&=0, 
\end{align}
\end{subequations}
where  $L_{\theta_i}$ and  $L_{\vp_i}$ are the extrapolation lengths for polar and azimuthal anchoring, respectively,  $L_{\theta_2}>L_{\theta_1}$ (the planar anchoring is stronger than the homeotropic one),  as before  we assume $\bar\theta_1=\pi/2$, $\bar\theta_2=0$ and the system of coordinates is chosen so that  $\bar \vp_1\equiv0$, and then the index is dropped by for $\bar\vp_2\equiv\bar\vp$.

\begin{figure}[ht]
\centering
\includegraphics[width=0.75\linewidth]{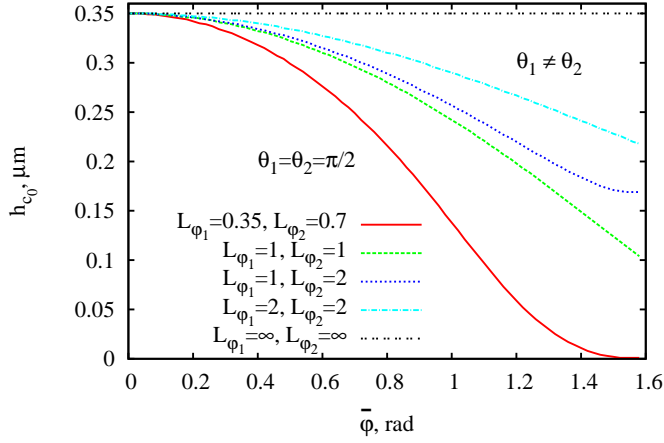}
\caption{The critical thickness \eqref{eq:hazim} between planar (or twisted) states with $\theta_1=\theta_2=\pi/2$ and HAN (or twisted-bent) states with $\theta_1\neq\theta_2$. As before we fix $L_{\theta_1}=0.35~\mu$m and $L_{\theta_2}=0.7~\mu$m, yielding $h_{c_0}=0.35~\mu$m for $\bar \vp=0$ or $L_{\vp_i}\to\infty$.}
\label{fig:phi2}
\end{figure}

The conventional uniform planar  and HAN ground states  can be found for a vanishingly small azimuthal anchoring $L_{\vp_i}\gg L_{\theta_i }$ and $\bar\vp=0,\pi/2$. Interestingly, if $\bar\vp\in(0,\pi/2)$ is an arbitrary angle, the boundary conditions \eqref{eq:4bc} are satisfied if and only if $\vp_1\neq\vp_2$ or $\theta_1=\theta_2=0$, irrespective of the smallness of the anchoring strength $W_{\vp_i}$. The planar state becomes linearly unstable with respect to the twisted state if $L_{\vp_2} - L_{\vp_1}<h<L_{\theta_2}-L_{\theta_1}$.
The states a) and c) become unstable with respect to the small perturbations of the polar angle $\theta$, resulting in states b) and d), respectively, if the thickness of the nematic film satisfies the following inequality 
\begin{multline} \label{eq:hazim}
h(L_{\vp_1}-L_{\theta_1}\sin^2\vp_1 )(L_{\vp_2}+L_{\theta_2}\sin^2(\vp_2-\bar\vp) )>   L_{\vp_1}L_{\vp_2}\times\\(L_{\theta_2}-L_{\theta_1})-L_{\theta_1}L_{\theta_2}(L_{\vp_2}\sin^2\vp_1 +L_{\vp_1}\sin^2(\vp_2-\bar\vp)).
\end{multline}
In the limit of the vanishing azimuthal anchoring \eqref{eq:hazim} yields the known result for the critical thickness $h_{c_0}\equiv L_{\theta_2}-L_{\theta_1}$~\cite{BB:1983}.  In presence of azimuthal anchoring the threshold $h_{c_0}$ cannot be defined explicitly, because of the dependence  of $\vp_1$ and $\vp_2$ on $h$ through the boundary conditions \eqref{eq:4bc}. In Fig.~\ref{fig:phi2} we plot the computed  threshold $h_{c_0}$, which has a smaller value than the Barbero--Barberi result, meaning that in presence of azimuthal anchoring the homogeneous planar (or twisted) states  are easily replaced by the HAN (or twisted-bent). Experimentally, he structure of the film with thickness 20-30~nm is unknown~\cite{ulysse:2009,ulysse:2010,cazabat:2011} and one cannot exclude the tendency of the director to form a twist configuration.


\begin{figure}[ht]
\centering
\raisebox{54mm}{a)}\includegraphics[height=5.5cm]{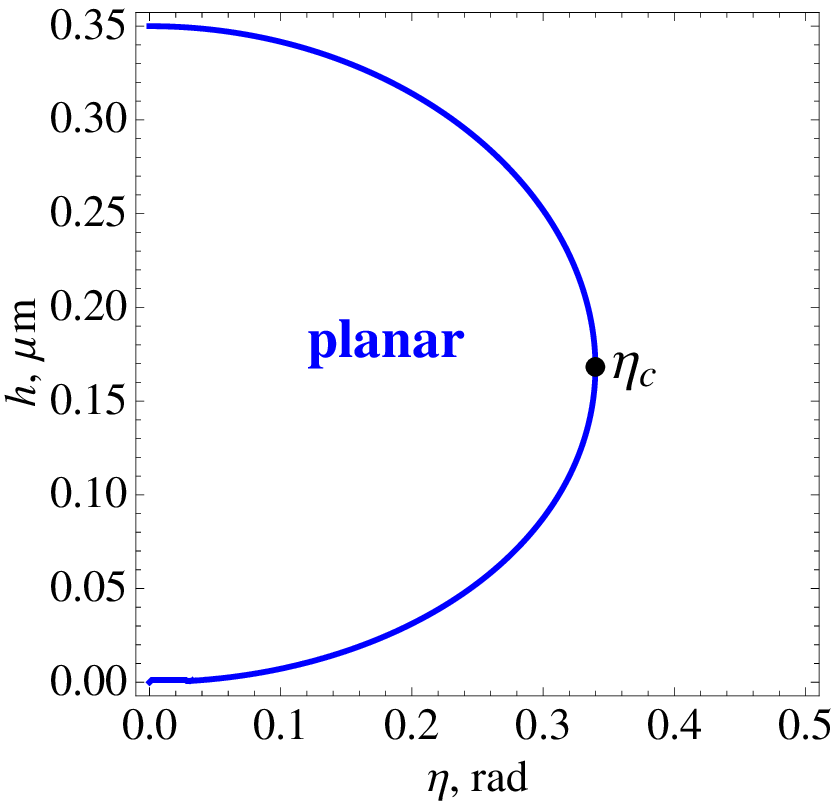}\hfil
\raisebox{54mm}{b)}\includegraphics[height=5.5cm]{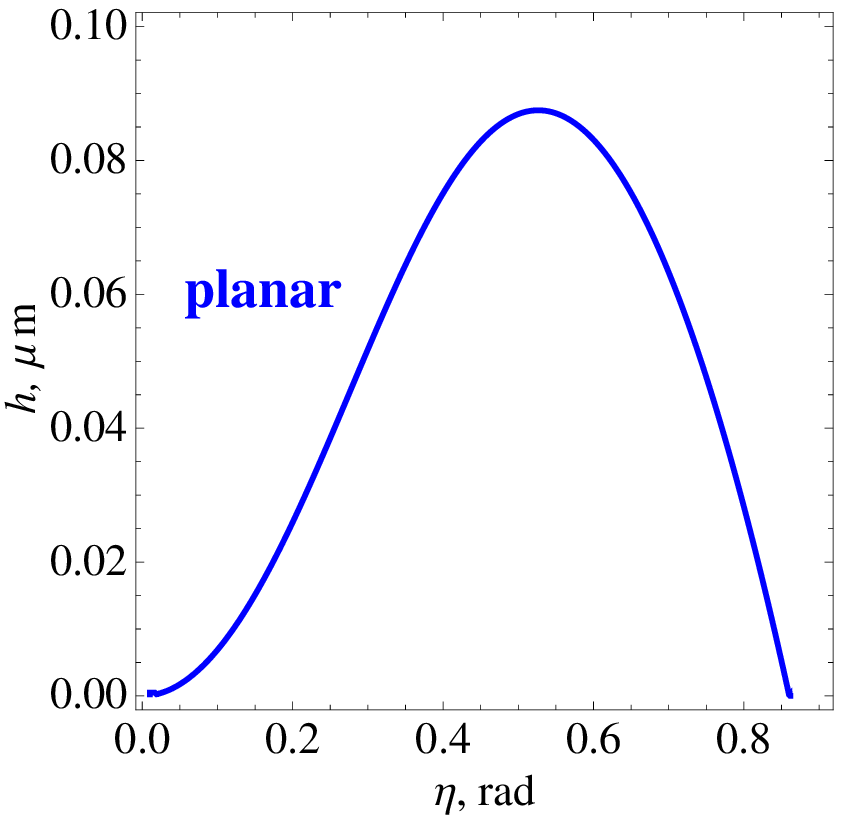}
\caption{The relative stability of the planar state w.r.t. perturbed state~\eqref{eq:hat} ($L_{\theta_1}=0.35~\mu$m, $L_{\theta_2}=0.7~\mu$m) a) $\ve^2$-term in~\eqref{eq:fhpm}  is always negative above  the critical value of the in-plane $\vp$-undulation  $\eta_c=\arcsin[(L_{\theta_2}-L_{\theta_1})/(L_{\theta_2}+L_{\theta_1})]$; b)~under the bell-shaped curve ${\cal F}<0$~\eqref{eq:fhpm}  the planar state is linearly unstable. We choose $\ve^2=0.5$, for smaller $\ve$ the peak of the curve shifts towards smaller $\eta$ and the area under the curve shrinks.}
\label{fig:hphi}
\end{figure} 

\subsection{Different scaling for $\theta$ and $\vp$}

Thin films with planar configurations  $\theta=\pi/2$ and $\vp=0$, can become unstable with respect to a small perturbations $\hat\theta$ and $\hat\vp$, admitting the following form
\be
\theta(\z)=\frac \pi2+\ve\hat\theta+\ldots,\qquad \vp(\z)=\hat\vp+\ve^2\hat \vp_1+\ldots
\ee
with $\ve$ being also small. Hence we have conjectured that the reorientation of $\bn$ in the plane of the film happens before  the change of the polar angle $\theta$. Indeeed, according to~\cite{jerome} as well as our analysis, the azimuthal anchoring should be one or two orders of magnitude smaller, resulting in energetically less expensive reorientation of the molecules in the plane of nematic--liquid substrate interface rather than change of the polar angle $\theta$.   Usually, $\hat\vp$ is considered to be negligible, resulting in the planar ground state and thus impossible coupling of two degrees of freedom. In our case, the solution of \eqref{eq:EL} up to $O(\ve)$ takes the form
\be\label{eq:hat}
\hat\vp(\z)=\eta \cdot\z,\qquad \hat\theta(\z)=A\cos(\eta\cdot \z) +B\sin(\eta \cdot\z),
\ee
with coupling of $\hat\vp$ and $\hat\theta$ resulting in negative contribution to the linearised  total free energy
\be\label{eq:fhpm}
{\cal F}=\frac K {2h}\bigg[\eta^2+\ve^2\bigg(\int_0^1 d\z\, \{(\p_\z\hat\theta)^2 -\eta^2 \hat\theta^2 \} +\frac{h}{L_{\theta_1}}\hat\theta^2(0)-\frac{h}{L_{\theta_2}}\hat\theta^2(1)\bigg)\bigg].
\ee
Although the first term is always positive the second term becomes negative  if the thickness of the film is $h<h_-$ or $h>h_+$, where
\be\label{eq:hpm}
h_\pm=\frac {\eta }{2 \sin\eta}\Big(\cos\eta (L_{\theta_2}-L_{\theta_1})\pm\sqrt{(L_{\theta_2}-L_{\theta_1})^2\cos^2\eta-4L_{\theta_1}L_{\theta_2}\sin^2\eta} \Big),
\ee
yielding  the Barbero--Barberi critical thickness $h_{c_0}=L_{\theta_2}-L_{\theta_1}$~\cite{BB:1983} in the limit of $\eta\to0$. The curve $h_{\pm}(\eta)$, shown in Fig.~\ref{fig:hphi}a, separates the region of the unperturbed planar state from the region where the instability towards the state with non-zero $\hat\vp$ and $\hat\theta$ may plausibly occur.  To quantify the destabilising effect of the second term compared to the stabilising effect of the first term we find the amplitude of the perturbation as described in the previous section~\ref{sect:ampl}. In Fig.~\ref{fig:hphi}b we plot the  curve when ${\cal F}=0$ so two competing effects cancel each other, giving the bell-shaped region for the planar state instability.

\section{Discussion and concluding remarks}

One of the main challenges in modelling thin nematic films is to describe the interplay of different length scales simultaneously~\cite{ulysse:2008,cazabat:2011}. The profile of nematic microdrop extends over micrometers while the thickness of the film experiences sudden jump from 3~nm to 30~nm. Similarly, the wavelength of observed stripes varies from microns till hundreds of microns while the thickness of the flat central part of nematic film changes in the range $20~{\rm nm}<h<0.5~\mu$m. For such small thickness the interfaces play a crucial role, and short-range interactions should be included in theoretical models.  This would require the development of  microscopic approaches, probably similar to~\cite{osipov:1997}, which would result in better understanding of  wetting and anchoring properties of nematic liquid crystals.  On the other side of the length scale, it is useful to account for the film profile by studying hydrodynamic (in)stabilities as described in~\cite{benamar:2001,cummings:2012}. Phenomenology is an alternative approach which  can be useful to capture both long-wavelength instability patterns as well as the structure of the ultrathin nematic film and to interprete experimental observations.

Here, within a  phenomenological framework, we have reconsidered the stripe instability in thin nematic films, subjected to antagonistic boundary conditions. The thickness of the film as well as azimuthal anchoring were treated as extrinsic parameters of our system related to a particular experiment, while the polar anchoring and the saddle--splay elasticity being intrinsic ones. Solving the variational problem within the one-constant approximation we identified the critical thickness  and the critical wavenumber, which determine the threshold of stripe instability. We (re)examined the effects of saddle-splay constant $K_{24}$ as well as  azimuthal anchoring on the critical thickness, wavelength and the amplitude of stripes at the bifurcation point. Then, we analysed the possibility of non-planar base state below the Barbero--Barberi critical thickness, assuming a certain {\it anzats} 
for  the biaxial order of nematic director in the vicinity of the interface or a certain scaling for the in-plane and out-of-plane degrees of freedom.  Considering different base states and their perturbations is relevant for predicting instability patterns such as stripes, squares and chevrons  as well as for establishing further connections with experiments. 

%
%


\section*{Acknowledgements}
It is our  pleasure to acknowledge the stimulating discussions with Anne-Marie Cazabat. O.\,V.\,M. was initially supported by the French National Research Agency (ANR), grant ANR-07-BLAN-0158.
%



\end{document}